%% file: main_revision.tex
\PassOptionsToPackage{table,xcdraw}{xcolor}
\documentclass[sigconf,nonacm]{acmart}

\AtBeginDocument{%
  }

\usepackage{multirow}
\usepackage{xcolor}
\usepackage{booktabs}
\usepackage{makecell}
\usepackage{color-edits}
\usepackage{enumitem}

\addauthor{lc}{orange}
\addauthor{wqy}{red}
\addauthor{zcw}{blue}

\begin{document}

\title{Promptile: Recovering and Reusing Prompting Strategies from LLM Interaction Histories}


\author{Qingyang Wan}
 \email{wanqy23@mails.tsinghua.edu.cn}
\affiliation{%
  \institution{Tsinghua University}
  \country{China}
}

\author{Chang Liu}
\email{cl2123@cam.ac.uk}
\affiliation{%
  \institution{University of Cambridge}
  \city{Cambridge}
  \country{United Kingdom}
}

\author{Zhoutong Ye}
\email{yezt24@mails.tsinghua.edu.cn}
\affiliation{%
  \institution{Tsinghua University}
  \country{China}
}

\author{Chengwen Zhang}
 \email{zcw25@mails.tsinghua.edu.cn}
\affiliation{%
  \institution{Tsinghua University}
  \country{China}
}

\author{Yanru Qian}
 \email{qyr25@mails.tsinghua.edu.cn}
\affiliation{%
  \institution{Tsinghua University}
  \country{China}
}

\author{Zhike Wang}
 \email{wangzhike233@ruc.edu.cn}
\affiliation{%
  \institution{Renmin University}
  \country{China}
}

\author{Jie Cai}
\email{jie.cai1@outlook.com}
\affiliation{%
  \institution{Nankai University}
  \city{Tianjin}
  \country{China}
}

\author{Chun Yu}
\email{chunyu@tsinghua.edu.cn}
\affiliation{%
  \institution{Tsinghua University}
  \city{Beijing}
  \country{China}
}

\author{Yuanchun Shi}
\email{shiyc@tsinghua.edu.cn}
\affiliation{%
  \institution{Tsinghua University}
  \city{Beijing}
  \country{China}
}
\renewcommand{\shortauthors}{Wan et al.}

\begin{abstract}
As people use large language models (LLMs) over time, they develop their own ways of prompting to achieve desired results. 
These useful experiences often remain scattered and unorganized across conversations. 
Yet existing personalization approaches primarily capture user preferences or explicitly defined workflows.
Through a formative study, we found that users develop prompting strategies that range from implicit instructions to explicit templates from these experiences. 
We further find that interaction behaviors provide evidence for these strategies, which can be decomposed and recombined for reuse across tasks.
Building on these findings, we present Promptile, a framework that identifies and validates users' prompting strategies from interaction histories. 
Promptile organizes these strategies into reusable components that can be retrieved, combined, and adapted to users' current tasks. 
Our evaluation shows that Promptile accurately recovers and reuses these strategies, better aligning model outputs with users' expectations while reducing repetitive prompting effort.
\end{abstract}





\maketitle

\section{Introduction}
\input{section-arxiv/10-introduction}
\section{Related Work}
\input{section/related-work.tex}

\section{Formative Study}
\input{section/formative-study.tex}
\section{System Design}
\input{section/system-design.tex}

\section{Technical Evaluation}
\input{section/technical-evaluation.tex}

\section{User Study}
\input{section/user-study.tex}

\section{Results}
\input{section/results.tex}

\section{Discussion}
\input{section/discussion.tex}

\section{Conclusion}
\input{section/conclusion.tex}




\bibliographystyle{ACM-Reference-Format}
\bibliography{references}

\appendix

\section{Further Analysis of Technical Evaluation Results}
\input{section/appendix-technical-evaluation.tex}

\section{User Study Participant Information}
\input{section/appendix-online-resources.tex}
\end{document}

%% file: section-arxiv/10-introduction.tex
As people repeatedly work with large language models (LLMs), they learn how to obtain outputs that better align with their intent. 
They do so by experimenting with prompts, refining their instructions, and adapting their requests in response to model outputs~\cite{zamfirescu2023johnny, Coscia2025OnGoal}.
Over time, some of these experiences become familiar ways of prompting that users consciously draw on when approaching similar tasks.

Prior work has explored several ways to support personalization and reuse in human-LLM interactions. 
Personalization systems draw on interaction histories to model user profiles~\cite{shaikh2025creating}, preferences~\cite{gao-etal-2026-beyond, wu-etal-2025-aligning}, memories~\cite{yen2024memolet, yuan-etal-2025-personalized, tan-etal-2025-prospect}, and task-specific objectives~\cite{ Lam_2026}, helping LLMs tailor their responses to individual users and contexts. 
Meanwhile, prompt and workflow management tools~\cite{bach-etal-2022-promptsource, chen2026from} and skill-based systems~\cite{yang2026autoskill} support reuse by preserving prompts, instructions, and workflows for future tasks. 
Together, these approaches enable LLMs to carry forward different forms of user and task knowledge, from inferred preferences and goals to explicitly represented procedures/answers.
However, users' prompting experience remains scattered across chat histories and is rarely used to guide future LLM interactions. 
This makes it difficult to inherit the ways users have learned to work with LLMs and adapt them to new tasks.

To understand how users' prompting experience could be recovered and reused, we analyzed 1,445 work-related prompts from nine participants and conducted retrospective interviews about their prompting practices. 
We found that users develop \textit{reusable prompting strategies} from their past experiences: ways of prompting that users have learned to associate with desired effects in recurring task contexts. 
Once developed, these strategies appear in users' interaction histories in different forms, ranging from scattered instructions to explicitly written templates.
Users' interaction histories also provide evidence of how these strategies develop and whether they remain useful. 
Specifically, users retain, refine, and reuse prompting approaches that produce desired results, while revising or abandoning those that do not.
Participants described retaining and reusing approaches they considered useful, and revising or abandoning them when outputs fell short of their expectations.
Furthermore, a single prompt often contains multiple strategies. 
Individual strategies can also be adapted and reused across different tasks.


Based on these findings, we designed and implemented Promptile, a system for identifying and reusing users' prompting strategies. 
Promptile first identifies recurring tasks in users' interaction histories and gathers the prompting methods and requirements expressed across related interactions. 
It then constructs reusable prompting strategies by organizing these elements around the task contexts and intended effects identified in our formative study.
To assess whether these strategies are supported by past interactions, Promptile examines model responses and subsequent user feedback when available. 
The resulting strategies are stored with their task contexts and supporting evidence, allowing both individual requirements and more complete workflows to be retrieved and reused in future tasks.


When users encounter a new task, Promptile retrieves potentially relevant strategies from their history and evaluates whether they are appropriate for the current task, taking into account the requested operation, task scope, and explicit instructions. This enables users to reuse previously established prompting approaches without restating their requirements or manually assembling prompt templates. Users can also inspect and adapt the retrieved strategies, selecting and combining them as needed for the task at hand. In doing so, Promptile aims to reduce the effort of repeatedly specifying how an LLM should perform familiar tasks, while allowing users to adjust their established approaches to new situations.


We evaluate Promptile using longitudinal LLM interaction histories from WildChat~\cite{wildchat}. We first examine whether Promptile can recover reusable prompting strategies from users' past interactions. We then evaluate whether these strategies can be appropriately retrieved and applied to held-out future tasks, improving model alignment with users' intents and desired outcomes. Finally, using real multi-turn task histories, we investigate whether recovered strategies help models satisfy users' requirements earlier, potentially reducing the need for repeated prompting and refinement.
Our results show that strategies recovered from prior interactions improved model alignment with users' intents and desired effects, as well as task performance.

Our work makes the following contributions:

\begin{enumerate}
    \item We empirically characterize \textbf{reusable prompting strategies} as units of users' accumulated prompting experience that connect an intent, a way of prompting the model, and a desired effect. Our formative study further characterizes how these strategies are developed, validated through interaction, reused, composed, and transferred across tasks.

    \item We present \textbf{Promptile}, a system and modeling approach for identifying, validating, storing, retrieving, and composing prompting strategies from heterogeneous interaction histories, making previously implicit prompting experience available as explicit reusable artifacts.

    \item We outline an offline evaluation using longitudinal interaction histories to assess strategy recovery, applicability to subsequent task instances, and requirement satisfaction in multi-turn tasks.
\end{enumerate}

%% file: section/related-work.tex
\subsection{User Modeling and Controllable Personalization}
User modeling has long served as a foundation for adaptive and personalized systems\cite{finin1986gums,fischer2001user}, which represent salient aspects of users so that systems can tailor content\cite{mysore2024pearl}, recommendations, and interaction accordingly\cite{kobsa2001generic}. This perspective has recently been extended to LLM-based systems, where personalization is often grounded in user profiles, interaction histories, authored content, and behavioral traces\cite{10.1145/3613904.3642212}. More recent work further broadens the scope of user modeling beyond task-local adaptation: systems such as GUM construct general user models from observations across everyday computer use\cite{shaikh2025creating}, while work such as InterQuest dynamically infers user interests during interaction rather than relying only on explicitly stated preferences\cite{mei2025interquest}. At the same time, prior HCI research has argued that effective personalization is not only about accurate inference, but also about giving users meaningful visibility and control over how adaptation happens\cite{kay2013creating,segijn2021literature}. However, most existing work models users at the level of broad traits, interests, goals, or histories while pays less attention to reusable prompting preferences as a distinct kind of user model: preferences that are expressed through interaction, recur across tasks, and can be deliberately reused in future prompting while remaining inspectable, revisable, and user-controlled over time.

\subsection{Memory, Personalization, and User Control in Cross-Session LLM Interaction}
Memory has become an increasingly important component of LLM-based systems\cite{guan2025survey} \cite{zhang2025survey,zhang2024personalization}. As LLM interactions extend beyond single sessions, supporting long-term memory has become increasingly important for maintaining continuity and personalization over time. Existing approaches to long-term LLM memory mainly rely on retrieval, summarization, or explicit memory updating\cite{liu2024memlong,chhikara2025mem0,tan2025prospect,yuan2025personalized}. Retrieval-based methods resurface relevant past conversations or user information\cite{packer2023memgpt}; summarization-based methods compress dialogue history into reusable representations\cite{wang2025recursively,liu2023think}; and memory-bank approaches maintain longer-term user models through update and forgetting mechanisms\cite{zhong2024memorybank}. Industry systems such as ChatGPT Memory reflect this trend by supporting persistent memories across chats with user controls for deletion, disabling, and temporary chat\cite{openai_memory_faq,google_gemini_memory}.
However, most existing memory mechanisms remain oriented toward recalling prior content—what the user said, what happened before, or what facts may be useful—rather than distilling reusable patterns in how users interact with LLMs\cite{yen2024memolet}. Even when they support personalization, they typically expose memory as retrieved conversation, summarized history, or stored facts, with only limited support for representing reusable prompting preferences as first-class, editable artifacts. At the same time, prior HCI work shows that users often have incomplete mental models of what AI systems remember and desire greater transparency and control over memory formation and reuse\cite{jones2025users,kay2013creating}. Promptile builds on this line of work by focusing specifically on prompt-related memory: not just remembering past prompts, but surfacing and managing reusable prompting preferences that users can inspect, revise, and carry across tasks.

\subsection{Prompting as Authored and Reusable Work}
Memory has become an increasingly important component of LLM-based systems \cite{guan2025survey,zhang2025survey,zhang2024personalization}. As LLM use extends beyond single sessions, long-term memory is increasingly needed to support continuity and personalization over time. Existing approaches mainly rely on retrieval, summarization, or explicit memory updating \cite{liu2024memlong,chhikara2025mem0,tan2025prospect,yuan2025personalized}. Retrieval-based methods resurface relevant past conversations or user information \cite{packer2023memgpt}; summarization-based methods compress interaction history into reusable representations \cite{wang2025recursively,liu2023think}; and memory-bank approaches maintain longer-term user models through update and forgetting mechanisms \cite{zhong2024memorybank}. Industry systems such as ChatGPT Memory reflect this trend by supporting persistent memory across chats with user controls such as deletion, disabling, and temporary chat \cite{openai_memory_faq,google_gemini_memory}.
However, most existing memory mechanisms focus on recalling prior content—what the user said, what happened before, or what information may be useful—rather than capturing reusable patterns in how users interact with LLMs \cite{yen2024memolet}. Even when they support personalization, memory is typically exposed as retrieved conversations, summarized history, or stored facts, with limited support for representing prompting preferences as first-class, editable artifacts. Prior HCI work further shows that users often have incomplete mental models of what AI systems remember and want greater transparency and control over memory formation and reuse \cite{jones2025users,kay2013creating}. Promptile builds on this line of work by focusing on prompt-related memory: not merely remembering past prompts, but surfacing reusable prompting preferences that users can inspect, revise, and carry across tasks.

%% file: section/formative-study.tex
To inform the design of our preference-extraction system, we conducted a formative study to understand whether and how reusable user preferences are expressed in prompts. In particular, we examined how such preferences can be identified, interpreted, and organized in ways that support later extraction and reuse. This study laid the groundwork for both a workflow for recognizing reusable preferences in prompts and a taxonomy for structuring them into reusable fragments.
\subsection{Qualitative Analysis}

\subsubsection{procedure}To evaluate prompt reusability and decompose preferences for system, we developed a structured taxonomy. We first conducted a formative study grounded in both prior literature and real user data. Prior work on prompt structure, components and patterns suggests that prompts often contain recurring elements related to user intent, response expectations, and reference use \cite{white2023prompt, mao2025prompts, jeoung2026promptprism}. 
Although existing frameworks offer valuable support for organizing prompt semantics, they are still ill-suited for accommodating the nuanced nature of user preferences. Consequently, these taxonomies lack the criteria to distinguish personal preferences from general task content, offering limited guidance on decomposing prompts into modular, storable units.



To fill the gap, we conducted an empirical analysis using real-world user prompts. We collected 1,445 historical prompts from nine experienced ChatGPT users. We then excluded 309 entries that were overly brief or lacked substantive content. Two researchers independently annotated preference-related sections using the initial taxonomy and resolved disagreements through discussion. After coding, we conducted a 15-minute semi-structured interview to understand participants’ original intentions behind the annotated prompts and their perspectives on extracting preferences for reuse.

\subsubsection{findings}
Our formative analysis revealed that user preferences are not isolated, static entities but are deeply embedded in conversational context and evolve through interaction. First, preferences are \textbf{highly contextual}. In particular, preferences expressed in follow-up refinements often rely on the preceding dialogue for clarity. For example, to understand what P1 meant by \textit{``make this simple enough for a primary school student''}, the preceding answer given by the chatbot is needed. Second, preferences exhibit a \textbf{performance-driven evolution} throughout the session. Participants often refine and adjust their instructions based on the quality of the LLM's output. We saw this in P2’s case, who went from \textit{``I want to present in a narrative way, so that even audiences unfamiliar with the topic can quickly grasp them clearly.''} to \textit{``storytelling expression for academic settings''} in their quest for an appropriate LLM response. We saw P2 consistently using this refined prompt in all subsequent interactions. Third, participants argued that a prompt only qualifies as a preference if it \textbf{consistently yields the desired results}. As P7 expressed: \textit{``If a prompt performs poorly, I don't think it’s something that should be saved... I would change my expression even if I believe the prompt bears what I want.''} This suggests that participants did not treat prompts as reusable preferences based on wording alone, but evaluated them through their demonstrated consistency in producing satisfactory outcomes.

Based on our formative study, we made 2 key observations:
\begin{itemize}[leftmargin=*]
 \item \textbf{Not all prompts contain reusable preferences.} The system needs a workflow for determining whether a prompt contains reusable preferences. It should take into account whether the prompt has been consistently successful and whether it has undergone refinement over time.
 \item \textbf{Conversational context is necessary when mining prompts.} The System must leverage the surrounding conversational history to resolve ambiguity in prompts.
\end{itemize}

\subsection{A Taxonomy of Prompt-Expressed User Preferences}

\newlength{\mycolwidth}
\setlength{\mycolwidth}{5.5cm}

\begin{table*}[]
\footnotesize
    \begin{tabular}{l|l|p{\mycolwidth}p{\mycolwidth}}
    \toprule
     Dimension&  What it captures& Definition & Example \\
    \midrule
    \multirow{3}{*}[0pt]{Context} 
        & General Domain
        & A brief characterization of the prompt's overall task type and application context.
        & Academic paper writing \\ \cline{2-4} 
        & Specific Topic
        & The key variable content in the prompt, normalized into standardized placeholders.
        & Application Essay Writing for [Balabala] University \\ \cline{2-4} 
        & Contextual information
        & Relevant information carried over from earlier turns in the same dialogue when the task depends on prior context.
        & The LLM generated an article for the user. The user now wants the LLM to \texttt{``expand on the "Improving the Path" section''}, which is one of the article's subtitles.\\ \cline{2-4} 
        & Input Description
        & Specifies what the provided input is, including its source, genre.
        & \begin{tabular}[t]{@{}p{\mycolwidth}@{}}This is a part of an HCI academic paper.\\ This is a section of an application essay.\end{tabular} \\ 
    \midrule
    \multirow{5}{*}[0pt]{Instruction} 
        & Task Instruction 
        & The explicit directives and goals that specify what the model should do.
        & \begin{tabular}[t]{@{}p{\mycolwidth}@{}}\texttt{`` Polish a piece of text.''}\\ \texttt{``Please analyze the language and fill in the missing semantics.''}\end{tabular} \\ \cline{2-4} 
        & Role / Persona 
     & The identity or expertise the model is asked to assume.
     & \texttt{``You are an HCI expert.''} \\ \cline{2-4} 
     & Scenario 
     & Who is speaking and under what context.
     & \texttt{``I am an eighty-year-old grandmother pursuing an HCI PHD. ''} \\ \cline{2-4} 
     & LLM Behavioral Constraint
     & Instructions that regulate how the model should interact with the user or proceed through the dialogue, rather than what content the response should contain.
     & \begin{tabular}[t]{@{}p{\mycolwidth}@{}}\texttt{`` When your opinion differs from mine, you need to persuade me.''} \\ \texttt{`` Please restate what you are going to do. If you understand my needs, then we can begin the discussion.''} \end{tabular} \\ \cline{2-4} 
     & Procedure Specification 
     & Instructs the model to follow an explicit sequence of steps or subtasks when producing response.
     & \texttt{`` You need to follow these steps: 1. ... 2. ... 3. ... ''} \\
    \midrule
    \multirow{3}{*}[0pt]{Supporting Info} 
    & Example 
    & User-provided input-output pairs or exemplars.
     & \texttt{``This is an example: ...''}  \\ \cline{2-4} 
     & Task-Relevant Background 
     & Provides supplementary information about the task background and character motivations, adding implicit knowledge needed to perform the task. 
     & \begin{tabular}[t]{@{}p{\mycolwidth}@{}}\texttt{`` Here is some knowledge for your reference and understanding: user study design should ...''}\\ \texttt{`` What I am currently doing is ...''}\end{tabular} \\ \cline{2-4} 
     & External Knowledge base 
     & Requires the model to use or follow objectively existing theories, data sources, etc. 
     & \texttt{``Based on ... theoretical tools ''} \\
    \midrule
    \multirow{2}{*}[0pt]{Output} 
        & Format 
        &  The structure, length, or organization of the response.
        & \begin{tabular}[t]{@{}p{\mycolwidth}@{}}\texttt{``No more than fifty words.''} \\ \texttt{``Write it as a paragraph.''} \end{tabular} \\ \cline{2-4} 
     & Style / Tone 
     & The linguistic style, tone, or register of the response.
     & \texttt{``Use simple language ''}\\
    \bottomrule
    \end{tabular}
    \caption{A taxonomy of prompt-expressed user preferences.}
\end{table*}

The taxonomy consists of four high-level dimensions: Context, Instruction, Supporting Information, and Output. Among them, Context captures information that helps interpret and match prompts (e.g., domain, input type), rather than being directly reused in subsequent prompts. Instruction, Supporting Information, and Output encode different types of reusable preferences that can be carried across interactions.

Table 2 presents the full taxonomy, including definitions and examples for each category. Specifically, Context captures situational cues including the General Domain, Specific Topic Contextual information and Input Description. Instruction specifies what the model should do and how it should act, including Task Instructions, Persona, LLM Behavioral Constraint ,and Procedure Specification. Supporting Information refers to Examples, Background Information, Reference Materials, and Knowledge Base required or provided for the model to use. Output include Format, Style or tone. Unlike Supporting Information, Context does not supply substantive content for task completion.
This taxonomy serves as the foundation for our system design. During interaction, the system retrieves and surfaces fragments based on contextual signals, enabling users to directly reuse and manipulate them.



%% file: section/system-design.tex
\begin{figure*}[]
  \centering
  \includegraphics[width=\textwidth]{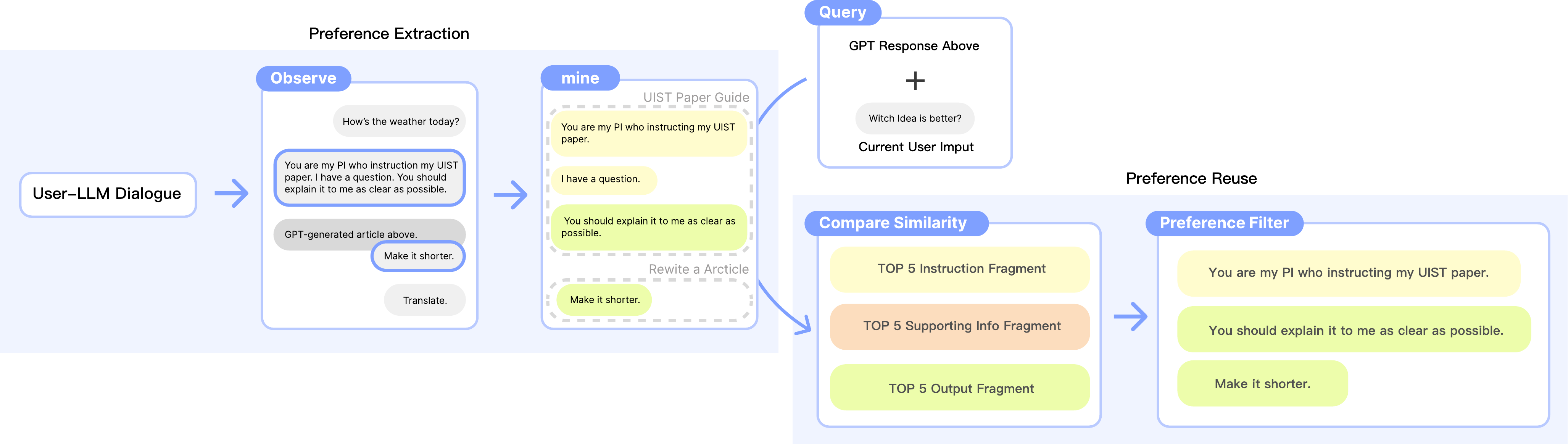}
  \caption{The Pipeline of Promptile System  }
  \Description{}
\end{figure*}

\begin{figure}[]
  \centering
  \includegraphics[width=0.5\textwidth]{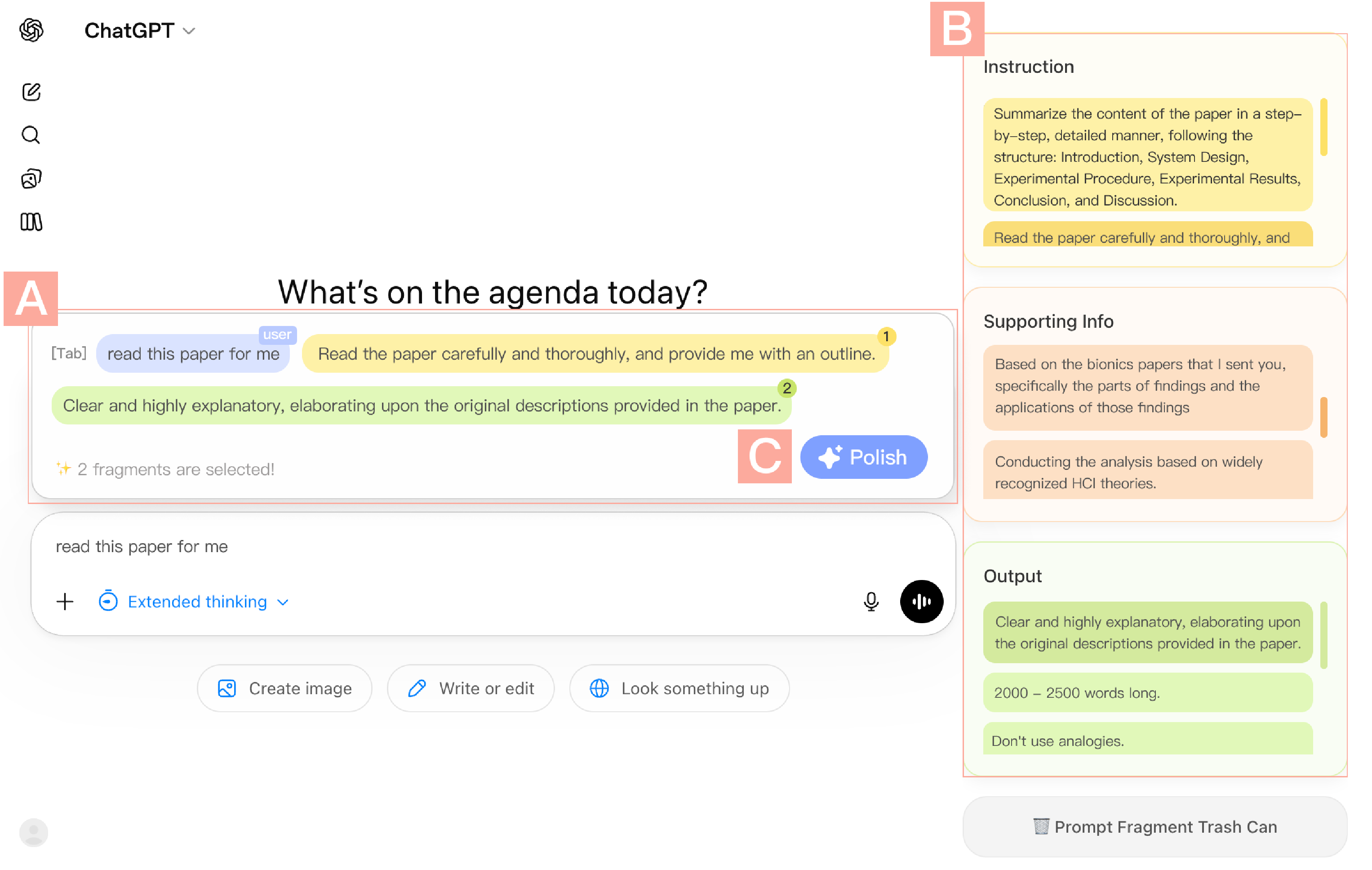}
  \caption{The User Inter Face of Promptile
  A. System input box. After filtering, the selected fragments are inserted into the system input box, where users can review how retrieved preferences are assembled into the current prompt context.
  B. Fragment panel. This panel displays retrieved fragments grouped by type (e.g., Instruction, Supporting Information, and Output). Users can interact with these fragments through drag-and-drop to add, remove, or rearrange them.
  C. Polish button. The Polish button reorganizes the selected fragments and adds lightweight transitions to make the system input more coherent while preserving the original content.
  }
  \Description{}
\end{figure}

Building on findings from Section 3, we designed Promptile, a system that aims to help users identify, extract, and organize reusable prompt preferences from their past interactions with LLMs.
The system design is organized around 4 components: chat history processing, context-aware retrieval and filtering, interactive fragment-based prompt assembly, and learning from user edits over time.

\subsection{Preference Fragment Extract from History}
The first module in Promptile transforms raw chat histories into reusable preference fragments that can support later retrieval and prompt composition. 
Given exported LLM chat history, Promptile first normalizes the history into a unified format to support downstream analysis. This step standardizes conversation records by aligning speakers and segmenting turns, while preserving the main meaning and structure of the interaction for later analysis. It then invokes two analysis agents: \textbf{Preference Observation} and \textbf{Preference Mining}.
The Preference Observation stage identifies valid, reusable user preferences from conversation histories. Specifically, it extracts actionable preferences that are contextually clear, proven effective, or refined over time. Consequently, it filters out noise like one-off task instructions and unstable or weak signals.
Promptile identified preferences into smaller reusable fragments Instead of storing an entire prompt or conversation as a monolithic artifact. The Preference Mining stage extract these preferences as fragments and categorized into semantic types in taxonomy.

The output of History Processing is therefore not a summarized conversation, but a structured preference repository composed of categorized prompt fragments. This provides the foundation for Promptile's functionalities: helping users recover, organize, and reuse their personal prompting preferences from past LLM interactions.
This design also reflects our view that preferences in prompts are better represented as modular and recombinable units than as fixed prompt templates.

\subsection{Context-Aware Retrieval and Filtering}

When a user composes a new prompt, Promptile uses the current draft (and the previous LLM responses in the same chat when available) to \textbf{retrieve reusable preference fragments from the user’s history-derived preference repository}. Retrieved fragments are presented on the right side of the interface, while a filtered subset is incorporated into the system input box as suggestions for the current task. 

To balance coverage and usability, Promptile first retrieves five candidate fragments from each of three categories—Instruction, Supporting Information, and Output—yielding 15 candidates in total. These candidates are then passed to an LLM-based fragment filtering agent, which removes redundant or inappropriate fragments and reranks the remaining ones for the current task context. 

Formally, For each candidate fragment $f_{ij}$ from historical prompt $i$, we compute a relevance score $S_{ij}$ by jointly considering the \textbf{current input}, the \textbf{current chat}, and the source preference item's \textbf{original context}:
\[
S_{his}=\frac{\mathrm{sim}(q_{chat}, f_{ij}) + \mathrm{sim}(q_{chat}, c_i)}{2}
\]
\[
S_{in}=\frac{\mathrm{sim}(q_{in}, f_{ij}) + \mathrm{sim}(q_{in}, c_i)}{2}
\]
\[
S_{ij} = 0.3 \cdot S_{his} + 0.7 \cdot S_{in},
\]

where $q_{chat}$ and $q_{in}$ are the embedding of the current chat context and current input, respectively, $c_i$ is the embedding of the \textit{Context} dimension (see Table \label{tab:preference-taxonomy} for the exact definition and examples) of preference item $i$, and $\mathrm{sim}(\cdot,\cdot)$ is cosine similarity function. All text embeddings are generated by a open-source text embedding model.

Before presentation, the system applies a lightweight deduplication step over retrieved candidates. Candidates are first sorted by relevance score. For each candidate fragment $c_i$, the system first checks \textbf{exact textual duplication}. If no exact duplicate is found, the system then performs \textbf{semantic deduplication} by comparing the fragment embedding $f_i$ against embeddings of other candidates. A candidate is removed if
\[
\mathrm{sim}(f_i,f_j)\ge \tau,
\]
We use $\tau=0.95$ in our implementation. This process reduces redundant suggestions while preserving the highest-ranked instance and merging provenance information from duplicates.

\subsection{Interactive Fragment Reuse and Prompt Assembly}


The interaction is designed to support simple, controllable reuse of past preference fragments during prompt writing. To let users quickly decide whether a past preference is worth reusing, the system shows fragments explicitly in the interface instead of silently inserting them into the prompt. To help users judge whether a fragment fits the current task, the system allows them to inspect its content and source. To give users flexible control over reuse, the system lets them insert, move, combine, and revise fragments as needed during prompt construction.


To support lightweight adjustment of which preferences to reuse, the interface provides a drag-and-drop workflow that lets users insert retrieved fragments directly into the system input box. Users can work from top-ranked candidate fragments and system-filtered selections, deciding which previously inferred preference units are worth reusing in the present task. This reduces both recall burden and typing effort. Once inserted, fragments remain editable and movable, so users are not locked into the system’s initial retrieval or arrangement.


Users may further refine retrieved fragments by deleting inserted ones, editing their wording, or reorganizing them in the prompt. Because even relevant retrieved preferences often require local adjustment, the system also provides an automatic \textit{Polish} function that reorganizes selected fragments into a more coherent and model-ready prompt while preserving their core content.


\subsection{Adapting from User Use and Edits}

To help the system better match users' needs over time, we treat their reuse and editing behaviors as feedback on the suggested fragments. The system records whether a fragment is accepted, rejected, replaced, or revised, and uses these signals to update future retrieval and recommendation priorities.

For example, when a user repeatedly rewrites a certain type of preference, the system retains the revised version and prioritizes it in similar future contexts. When a fragment is often rejected or removed in a particular situation, the system lowers its priority in comparable cases. In this way, the system gradually adapts preference representations based on how users actually reuse and modify them, rather than simply storing static fragments from past prompts.

This design turns preference reuse into an ongoing feedback loop: automation helps surface reusable preferences, while user interactions help refine how those preferences are represented and retrieved over time.

\subsection{Implementation Details.}
Promptile is implemented in Python with a FastAPI backend. We use Azure OpenAI's gpt-5.1 endpoint for preference analysis, fragment filtering, polishing, and prompt assembly. We use Qwen3-Embedding-0.6B for text embedding. Preference fragments are cached at the fragment level, ranked by cosine similarity to the current input and interaction context, semantically deduplicated, and then filtered by an LLM-based reranking step before being assembled into the final prompt draft.

%% file: section/technical-evaluation.tex
{\bfseries Evaluation Setup.} We recruited 9 participants to assess the agent pipeline. None of the prompts used in this evaluation had been analyzed in the earlier formative study. On a 7-point Likert scale, participants rated three aspects: (1)~whether the pipeline correctly determined whether a prompt contained reusable preferences; (2)~whether the pipeline's reasoning for that judgment was accurate; and (3)~whether the extracted preferences were comprehensive and correctly classified into the taxonomy dimensions.

{\bfseries Results.} The pipeline achieved high performance on preference identification, with a mean accuracy of 91\% (SD = 12\%). Participants also rated the pipeline's reasoning about preference presence highly (M = 6.49, SD = 0.50), indicating that its judgments were generally straightforward and credible. Ratings for individual dimensions were also strong overall: Context (M = 6.43, SD = 0.42), Instruction (M = 6.63, SD = 0.31), Supporting Information (M = 6.10, SD = 0.86), and Output (M = 6.42, SD = 0.53). Overall, these results suggest that the pipeline can provide sufficiently reliable structured preference representations for downstream reuse. For the rare bad cases, we present examples and further analysis in Appendix \ref{sec:appendix-technical-eval}.


\begin{figure}[htbp]
    \centering
    \includegraphics[width=0.4\textwidth]{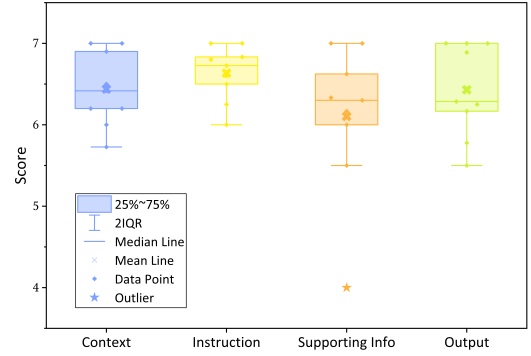}
    \caption{Technical evaluation results for preference extraction. \textnormal{Boxplots show participant ratings (N=9) on a 7-point Likert scale assessing whether the extracted preferences were comprehensive and correctly classified into the taxonomy dimensions}}
    \label{fig:interface}
    \Description{}
\end{figure}

\begin{table}[htbp]
\resizebox{0.6\linewidth}{!}
{
\begin{tabular}{ccc}
\toprule
& Accuracy & Score \\
\midrule
Correct& \multirow{2}{*}{$91.2\% \pm 11.9\%$} & $6.98 \pm 0.03$ \\
\cline{1-1} \cline{3-3} 
Incorrect& & $3.65 \pm 1.73$ \\
\bottomrule
\end{tabular}
}
\caption{Performance of the preference identification pipeline. \textnormal{The table shows the accuracy of determining reusable preferences and the participant ratings for the pipeline's reasoning in correct versus incorrect cases.}}
\end{table}

%% file: section/user-study.tex
To evaluate the effectiveness of our system, we conducted a within-subject user study comparing it against participants’ regular ChatGPT workflow with global memory. This study focused on the following three research questions:

\textbf{RQ1.} Can Promptile reduce the effort of writing prompts?

\textbf{RQ2.} Can Promptile make preference reuse more understandable and controllable?

\textbf{RQ3.} How do users use and experience Promptile when writing prompts?



\subsection{Study Design}
The study used a within-subject design with two conditions. Participants completed two matched paper-reading tasks under two conditions: \textit{System} and \textit{Baseline}. We counterbalanced both condition order and task materials across participants.

In the \textit{System} condition, participants sed our Chrome extension for preference reuse. The extension retrieved reusable preference fragments from each participant’s own historical prompts and allowed them to inspect, insert, edit, reorder, and refine these fragments during prompt construction.

In the \textit{Baseline} condition, participants used standard ChatGPT-5.3 through their own personal accounts, with ChatGPT’s global memory enabled for all participants, and followed their usual interaction workflow. No Thinking mode or additional interface support was provided in this condition.

Each participant completed two tasks, one in each condition. The tasks were designed to be as similar as possible in structure and difficulty while differing in source material. To reduce order effects, we counterbalanced both condition order and task materials. The 16 participants were divided into four groups to balance system exposure and paper assignment.

\subsection{Participants}
We recruited participants through a screening questionnaire completed by 43 respondents and selected 16 for the formal study. All participants were students (13 female, 3 male), aged 20–31 years (M=22.43, SD=3.10). They reported frequent LLM use, averaging 6.7 days per week (SD=1.03), and rated their familiarity with AI at 5.14 on a 7-point scale (SD=1.16). We prioritized participants with substantial LLM experience whose prior usage included reading academic papers. We limited recruitment to students because historical prompts from working professionals often involve job-related content constrained by non-disclosure agreements, making them unsuitable for use in our study.

\subsection{Tasks}
The study tasks involved reading two humanities papers, both about the same well-known novel but with different analytical focuses. The tasks were designed to be as similar as possible in structure and difficulty while varying in source material. Participants were asked to complete each task within 15 minutes using their usual reading and prompting habits. We intentionally chose a task domain that was already common in participants’ real-world LLM use. This improved ecological validity while creating a setting in which prior prompting preferences remained relevant, but still required adaptation rather than direct reuse.

\subsection{Study Procedure}
\subsubsection{Chat History Collection}
Before the study, we asked each participant to export their own LLM chat history. After anonymization, we retained prompts from approximately the most recent month of their history, resulting in a personal retrieval pool of 192 prompts on average per participant (min = 41, max = 594, SD = 173.4062187). We used these prompts as the source material for preference retrieval.

\subsubsection{Session Procedure}
At the beginning of each session, participants were introduced to the study and instructed to complete the tasks according to their usual paper-reading and prompting habits. They were also asked to think aloud throughout the session so that we could capture their reasoning, uncertainties, and reuse strategies during prompt construction.

Each participant then completed two assigned tasks, one under each condition. Each task lasted 15 minutes. After completing each task, participants filled out the same post-task questionnaire. At the end of the session, they also completed a final Likert-scale questionnaire and participated in a semi-structured interview. In total, each session lasted approximately one hour, and participants received \$20 in compensation.

\subsection{Measures}
We assessed the system using subjective, behavioral, and interaction-based measures. Subjectively, we used NASA-TLX \cite{hart1988development} for perceived workload, UMUX-Lite \cite{10.1145/2470654.2481287} for perceived usability. In addition, we included several study-specific Likert-scale items to capture constructs not adequately covered by standardized workload and usability measures\cite{hornbaek2006current}. For metrics that passed the Shapiro-Wilk normality test, we applied paired t-test for statistical significance. For other metrics, the Wilcoxon signed-rank test was used.

Rely on prior researches\cite{10.1145/3490493}, we used these items to probe agency and perceived system performance. 
as well as helpfulness for preference reuse, sense of control, transparency, and trust. These constructs have been identified as important in prior work on human–AI interaction, especially for AI systems that expose intermediate materials, memory, or prompting structures for user inspection and manipulation. Behaviorally, we recorded task completion time, number of prompt turns, number of submitted prompts, and characteristics of the final prompt to capture efficiency and prompting effort across conditions. In the System condition, we also logged fragment-level interactions, including how many fragments were retrieved, selected, edited, deleted, added, or polished, to understand how participants engaged with reusable preference fragments in practice.


%% file: section/results.tex
\begin{figure*}[htbp]
  \centering
  \includegraphics[width=\textwidth]{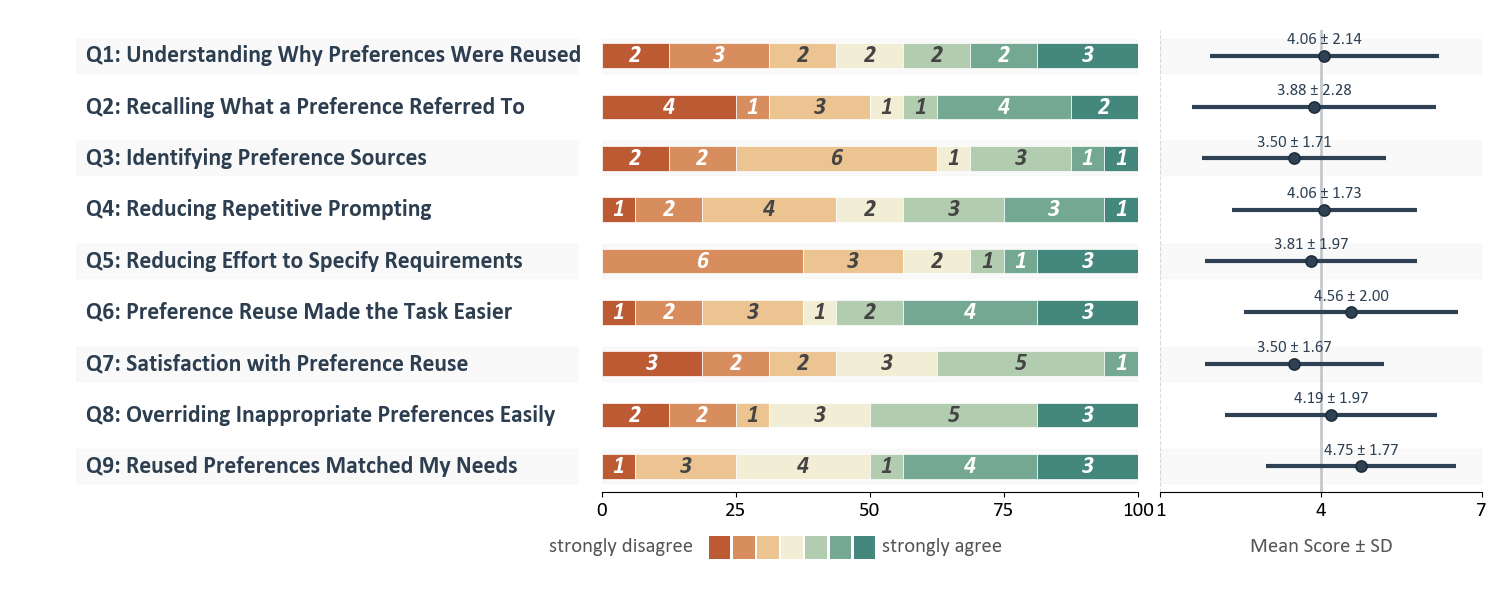}
  \vspace{-2em}
  \caption{\textbf{Participants’ Likert-scale ratings (1–7) for nine statements about preference reuse.} \textnormal{The stacked bars show response distributions across scale points (numbers indicate counts), and the right panel reports each item’s mean with standard deviation (Mean ± SD). Overall, participants expressed positive perceptions of preference reuse, with higher ratings on perceived task support and need matching, and comparatively lower ratings on recalling what a preference referred to and identifying preference sources.}}
  \Description{.}
\end{figure*}

\subsection{Promptile is an Effective Tool}

\begin{table}[htbp]
\centering
\caption{Comprehensive Comparison of Behavioral and Subjective Measures (N=16) }
\label{tab:comprehensive_evaluation}
\begin{tabular}{lcc}
\toprule
\textbf{Measure} & \textbf{Promptile} & \textbf{ChatGPT} \\
\midrule

Manual Input Amount (chars) & $\mathbf{51.38 \pm 37.09}$ & $91.63 \pm 51.60$ \\
Final Prompt Length (chars) & $\mathbf{205.76 \pm 108.44}$ & $176.81 \pm 66.98$ \\
Interaction Turns (count) & $\mathbf{1.88 \pm 0.72}$ & $2.69 \pm 1.01$ \\
Average Input Time (sec) & $145.93 \pm 95.48$ & $137.21 \pm 65.69$ \\
\bottomrule
\end{tabular}
\end{table}

\begin{figure*}
    \centering
    \includegraphics[width=\textwidth]{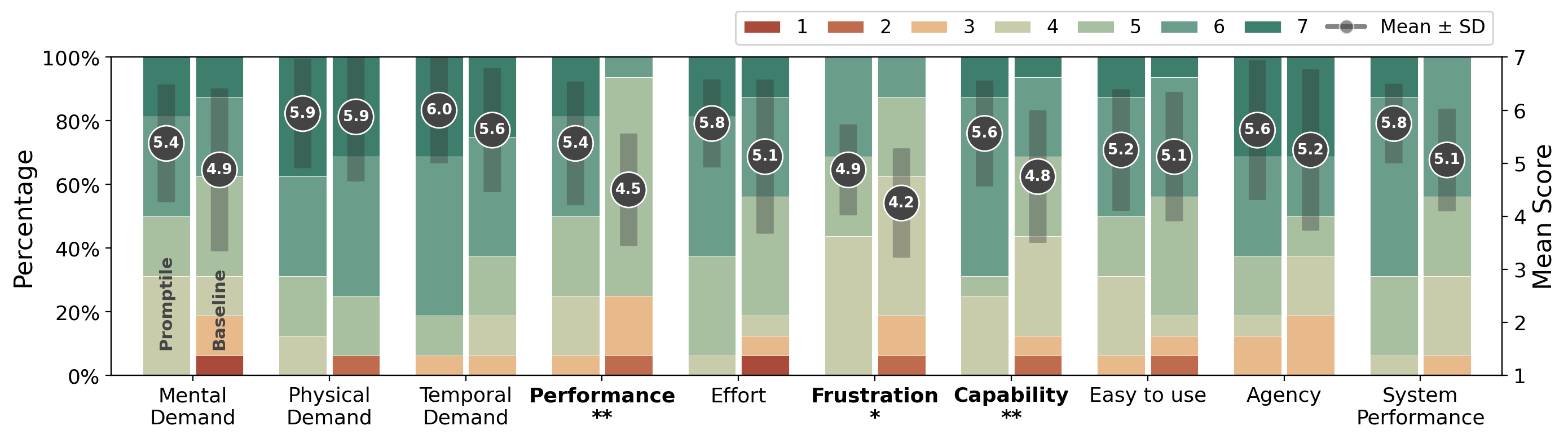}
    \caption{Results of the comprehensive subjective evaluation (N=16). \textnormal{The chart compares Promptile and the Baseline across NASA-TLX workload dimensions, system usability, and study-specific user experience metrics on a 7-point Likert scale. Note that standard NASA-TLX scales are reversed here: higher scores correspond to lower workload. Asterisks denote statistical significance (* p < 0.05, ** p < 0.01).}}
    \label{fig:interface}
    \Description{}
\end{figure*}

\subsubsection{Less Burden, Better Prompts}
Compared with the Baseline, Promptile significantly reduced the amount of manual input required per task (Promptile: M = 51.38, SD = 37.09 vs. Baseline: M = 91.63, SD = 51.60; p = 0.015). At the same time, final prompts produced with Promptile were longer on average (Promptile: M = 205.76, SD = 108.44 vs. Baseline: M = 176.81, SD = 66.98), suggesting that users were able to construct more complete instructions while typing less, although this difference did not reach statistical significance. Promptile also tended to reduce the number of interaction turns required to complete a task (Promptile: M = 1.88, SD = 0.72 vs. Baseline: M = 2.69, SD = 1.01). This directional pattern was also reflected at the individual level: 7 participants required fewer turns with Promptile, 2 required more turns, and 7 showed no change. In addition, participants reused an average of 4.85 fragments per task iteration (SD = 4.37), indicating that retrieved preferences were actively incorporated into prompt construction. Taken together, these results suggest that Promptile reduced the effort of composing prompts from scratch while supporting users in assembling richer instructions earlier in the interaction.

Participants felt the system was compelling, as it enhanced both task utility and their writing efficiency. The participants reported significantly better perceived self performance with Promptile than with the Baseline (system = 5.38 vs. Baseline = 4.50, p = 0.003) and felt that the system’s capabilities met their needs significantly better (system = 5.56 vs. Baseline = 4.75). Frustration is also significantly lower in the task (system =2.1 vs. Baseline=2.75). In the interviews, participants described that Promptile reduced the burden of prompt crafting (P3, P12), helped them compose a better prompt (P1, P4), and reminded them of useful prompt formulations they might otherwise have forgotten (P1). Taken together, these findings suggest that Promptile supported not only more efficient prompt crafting, but also more comprehensive articulation of user requirements.

\subsubsection{Promptile is useful, explainable, and controllable}

The results suggested that participants experienced Promptile’s reuse mechanism as transparent, helpful, and easy to control. On a 7-point scale, participants generally understood why particular preferences were reused in the current task (M = 5.44), could recall what those preferences referred to (M = 5.56), and could tell where reused preferences came from (M = 5.56). Together, these responses suggest that Promptile made reuse transparent. As P8 noted in study, \textit{``I knew at a glance that this was something I had written.''}

Participants also viewed preference reuse as useful for reducing effort. They agreed that reuse reduced repetitive prompting (M = 5.63), reduced the effort needed to describe their requirements (M = 5.44), and made the task easier overall (M = 5.50). P1 said \textit{``The fragment \texttt{`use literature review methods'} provided by Promptile led to the most compelling answer.''} Meanwhile, P3 noted that \textit{``I can use less effort to get GPT to accomplish what I want.''} At the same time, users reported strong agreement that reused preferences could be modified, removed, or overridden when needed (M = 6.13), indicating that the system preserved user control rather than imposing prior preferences rigidly. Consistent with this pattern, participants(P4, P5, P6, P9, P14, P15) also described reused fragments as easy to inspect and adjust. For example, P15 said \textit{``Deleting (fragments) felt really satisfying.''}. P9 also expressed that \textit{``The interaction felt very smooth overall. Removing, editing, and overriding all worked well and felt intuitive.''}

This combination of perceived usefulness and controllability is important. It suggests that participants preferred Promptile not because the suggestions were always directly applicable, but because they were transparent and easy to revise. P16 said: \textit{``Some don't really fit (the current task), so I edited them.''} Consistent with this interpretation, participants were generally satisfied with how the system reused their preferences (M = 5.00) and felt that the reused content broadly matched their needs in the task (M = 5.25). Overall, these responses suggest that Promptile’s benefits came not only from reducing burden, but also from making reuse transparent and keeping users in control.

\subsection{From Writing Prompts to Curating Preferences}
\subsubsection{Building prompts by pruning and revising retrieved fragments}

Promptile changed how participants constructed prompts. As reported in Table \ref{tab:comprehensive_evaluation}, Promptile reduced the amount of the number of characters typed manually by the participant, excluding reused fragments required per task, and tended to reduce the number of interaction turns, while average input time was descriptively longer under Promptile than under the Baseline (Promptile: M = 145.93, SD = 95.48 vs. Baseline: M = 137.21, SD = 65.69). This pattern places more of the observed prompt work before submission, during prompt assembly and adjustment, rather than only after submission through additional rounds of chat.

Interaction logs show that participants frequently modified retrieved material during composition. Fifteen of sixteen participants changed at least one suggested fragment during use. Across task sessions, participants reused an average of 4.85 retrieved fragments per task iteration (SD = 4.37). They also deleted retrieved fragments an average of 1.24 times per task (SD=1.41), directly revised retrieved fragments 0.32 times per task(SD=0.43), and added new fragments 0.19 times per task (SD=0.41). Here, ``deletion'' refers to  removing an entire retrieved fragment , ``revision'' refers to add, edit a fragment , and ``addition'' refers to newly drag a created fragments in to input box. These counts show that participants commonly worked from retrieved suggestions and then adjusted them by removing or revising material that did not fit the current task.

\subsubsection{Promptile as a collaborative aid for prompt construction}

Participants did not see Promptile as automatically generating prompts for them, but as actively supporting their own prompt construction process. Retrieved suggestions helped narrow possible formulations, guide further refinement, prevent omissions, and resurface previously used expressions or preferences. As P2 noted, \textit{``the suggestions helped narrow the range and indicate which direction to adjust the prompt''}. Similarly, P6 said \textit{``the system helped prevent missing important points''}, while P10 and P14 described it as \textit{``recalling prior phrasing and reminding them of useful additions''}. Together, these accounts suggest that \textbf{participants understood Promptile not as replacing their own efforts and agency, but as assisting them by contributing candidate formulations and reusable preference material during composition.}

Some participants further said that this process reduced the pressure of having to phrase everything precisely on their own. Instead of starting from scratch, they could respond to suggestions, inspect retrieved material, and decide what to keep or revise. As P3 and P9 described, this made the interaction feel less stressful and more collaborative, as if they were working together with the system. These accounts suggest that participants experienced Promptile as supporting a more collaborative form of prompt composition that involved curating system-provided material rather than meticulously writing every instruction from scratch.

%% file: section/discussion.tex
\subsection{Why Spending More Time Leads to \textbf{\textit{Less}} Frustration}

Average input time was descriptively longer with Promptile, while participants reported less frustration and a better experience than in the Baseline. We think this happened because Promptile changed the nature of the work rather than just adding more of it. In the Baseline, users often had to fix their prompts repeatedly after seeing a poor response. P16 described this as \textit{``a frustrating back and forth process that often felt like a long conversation without a clear result''}.

Promptile changes this by showing prompt materials as visible fragments on the screen \cite{scaife1996external, hollan2000distributed}. This allows users to recognize and choose useful parts instead of trying to remember and write everything from scratch. These fragments act as editable objects that users can adjust or combine easily \cite{shneiderman1997direct}. While it takes more time to review and select these fragments before submitting the prompt, this work is more manageable and less frustrating than trying to fix a prompt after an unsatisfactory response.

The extra time spent with Promptile was used for manageable and well-directed tasks like choosing and editing suggestions, which gave users more control. This made the process feel more like editing and organizing rather than write in a blank input box. This suggests that the \textbf{type of effort involved is more important for user experience than the total time spent}. Therefore, Promptile's effectiveness can be partly attributed to the \textbf{shift from \textit{writing prompts} to \textit{curating preferences}.}

\subsection{A Paradigm Shift for Personalizing LLMs}

The results from the previous section show that Promptile represents a paradigm shift from prompt writing to preference curation. This shift goes beyond one-time interactions. When interacting with Promptile, participants did not treat prompts merely as transient inputs for completing the current task. They also treated parts of prior prompts as worth preserving and resurfacing later, including preferred phrasings, recurring requirements, desired output structures, and, in some cases, useful supporting materials. This suggests that some of the value users place in prompting lies not only in specifying an immediate request, but also in externalizing ways of working with the model that remain useful over time.

Promptile made this longer-term layer of prompting more visible. By surfacing prior prompt fragments as reusable candidates, the system exposed how prompt content can function as a form of personal resource rather than only as disposable task input. Prompting now involves building up a repertoire of reusable ways of asking, constraining, and evaluating model behavior. This perspective helps explain why participants often valued retrieved fragments even when they did not reuse them verbatim: the benefit was not simply recovering old text, but recovering previously articulated preferences that could be reconsidered in the present situation. The usefulness of reuse therefore did not come from exact repetition alone. It came from making prior preferences available in a form that users could actively reinterpret, reframe, and adapt. In some sense, Promptile served as a form of steerable and malleable memory readily available to users.

Viewing preferences as a form of malleable memory sheds light on how to improve the design of personalized LLM systems. Many personalization mechanisms operate primarily in the background: they store user information, update profiles, and adapt outputs with limited visibility into what has been remembered or how it is being applied. Our findings suggest that this is only part of what support for personalization should entail. When reusable preferences are surfaced as visible and editable interaction material, users can better inspect what the system is drawing on, judge whether it fits the current context, and decide what should be kept, revised, or discarded. Supporting reuse, then, is not only a matter of retrieval accuracy. It is also a matter of making prior user expressions legible and controllable at the point of use.

%% file: section/conclusion.tex
In conclusion, this work highlights the importance of making long-term memory in personalized LLM systems more structured, transparent, and user-controllable. By introducing Promptile, we demonstrate how user preferences can be systematically captured, organized, and reused through a four-dimensional taxonomy, transforming them into modular and explainable prompt fragments. This approach not only reduces repetitive prompting effort but also improves output quality and enhances users’ sense of agency in interacting with LLMs. Our findings suggest that shifting from implicit, recall-based prompting toward explicit, editable preference curation enables more efficient and adaptable human–AI collaboration. By allowing users to inspect, modify, and recombine their preferences over time, Promptile fosters a more dynamic and user-in-the-loop memory paradigm.

%% file: section/appendix-technical-evaluation.tex
\label{sec:appendix-technical-eval}
The technical evaluation showed that our preference extraction pipeline performed well overall, and that the taxonomy of preferences in prompts mostly aligned with the users' understanding. However, we did encounter a few bad cases. These cases mainly occurred when prompts were noisy or underspecified. Some participants used fragmented phrases, typos, or minimal wording, which made both preference identification and taxonomy classification more difficult. Participants also noted that the pipeline occasionally appeared to infer implicit preferences from context despite being instructed to extract only explicit ones. The examples are as follows: 
\begin{quote}
\hangindent=1.5em
\hangafter=1
\texttt{``
...Based on the paper, identify the factors that affect consumers’ donation or purchase behavior,
preferably those that can be quantified. Then summarize how each factor is defined and measured in the paper.''~-P3
}
\end{quote}

\noindent\hspace*{1.5em}The pipeline inferred that:

\noindent\hspace*{1.5em}{\bfseries Output:} ~\texttt{{\bfseries List the factors one by one.} For each factor, include: the factor name, its conceptual definition in the paper, and how it is measured.}

\begin{quote}
\hangindent=1.5em
\hangafter=1
\texttt{``
...Help me summarize the core disputes and foreign-related elements in these two cases.''~-P4
}
\end{quote}
\noindent\hspace*{1.5em}The pipeline inferred that:

\noindent\hspace*{1.5em}{\bfseries Output:} ~\texttt{Summarize the cases separately. For each case, list {\bfseries in parallel} its core dispute and foreign-related elements.}

These cases suggest that the root cause of these bad cases is not flaws in the taxonomy, but the challenges posed by short and ambiguous prompts, as well as the sometimes vague boundary between explicit and implicit expressions of preference. Since these bad cases are relatively few in number and do not hinder the system in a significant way, we leave handling them to future work.

%% file: section/appendix-online-resources.tex
\begin{table}[htbp]
\caption{Participant information for the user study.}
\label{tab:user-study-participants}
\resizebox{\linewidth}{!}
{
\begin{tabular}{cclcccc}
\hline
\multicolumn{1}{c}{\textbf{ID}} & 
\multicolumn{1}{c}{\textbf{Age}} & 
\multicolumn{1}{c}{\textbf{Major}} & 
\multicolumn{1}{c}{\textbf{Gender}} & 
\multicolumn{1}{c}{\makecell[c]{\textbf{Weekly LLM} \\ \textbf{Usage (Days)}}} & 
\multicolumn{1}{c}{\makecell[c]{\textbf{LLM} \\ \textbf{Literacy}}} & 
\multicolumn{1}{c}{\makecell[c]{\textbf{Monthly} \\ \textbf{Prompts} }} \\
\hline
p1 & 20 & Design & Female & 7 & 5 & 145 \\
p2 & 25 & Design & Female & 7 & 6 & 150 \\
p3 & 21 & Design & Female & 7 & 6 & 297 \\
p4 & 20 & Japanese & Female & 7 & 5 & 89 \\
p5 & 20 & Industrial Engineering & Female & 7 & 6 & 209 \\
p6 & 20 & Automation & Female & 7 & 6 & 594 \\
p7 & 21 & Architecture & Female & 7 & 5 & 118 \\
p8 & 23 & Digital Media & Female & 7 & 5 & 57 \\
p9 & 21 & Architectural Environment & Female & 7 & 5 & 59 \\
p10 & 22 & Language and Literature & Female & 7 & 4 & 67 \\
p11 & 25 & Environmental Studies & Female & 7 & 4 & 339 \\
p12 & 25 & Medicine & Male & 7 & 4 & 64 \\
p13 & 19 & History & Female & 3 & 3 & 239 \\
p14 & 21 & Law & Female & 5 & 4 & 41 \\
p15 & 25 & Computer Science & Male & 7 & 7 & 68 \\
p16 & 31 & Computer Science & Male & 7 & 7 & 550 \\
\hline
\end{tabular}
}
\end{table}